\documentclass[12pt]{article}
\usepackage{amsmath}
\usepackage{epsf}
\usepackage{epsfig}
\usepackage{graphicx}
\usepackage[dvips]{lscape}
\usepackage{rotating}
\usepackage{subeqnarray}
\usepackage{setspace}
\usepackage{natbib}
\usepackage{afterpage}

\begin{document}
\bibliographystyle{stdbib}

\title{~~
~~ \\[-1in]
 {\Huge \bf Is It Real, or Is It Randomized?: \\ [.25in] A Financial 
Turing Test\thanks{The views and opinions expressed in this article are 
those of the authors only, and do not necessarily represent the views and 
opinions of AlphaSimplex Group, MIT, Northeastern University, or any of 
their affiliates and employees.  The authors makes no representations or 
warranty, either expressed or implied, as to the accuracy or completeness 
of the information contained in this article, nor are they recommending 
that this article serve as the basis for any investment decision---this 
article is for information purposes only.  Research support from 
Northeastern University and MIT Lab for Financial Engineering is 
gratefully acknowledged. We thank Dolphy Fernandes for programming 
assistance.}}} 

\author{\\[.0675in] \textbf{Jasmina\ Hasanhodzic,}\thanks{Senior Research Scientist,
AlphaSimplex Group, LLC, One Cambridge Center, Cambridge, MA 02142,
(617) 475--7106 (voice), {\tt jasmina@alum.mit.edu} (email).}
\textbf{ Andrew W.\ Lo,}\thanks{Harris \& Harris Group Professor, MIT
Sloan School of Management, and Chief Investment Strategist,
AlphaSimplex Group, LLC, MIT Sloan School of Management, 50 Memorial
Drive, E52--454, Cambridge, MA 02142,  (617) 253--0920 (voice), {\tt
alo@mit.edu} (email).} \textbf{ and Emanuele\ Viola}\thanks{Assistant
Professor, Northeastern University, College of Computer and
Information Science. Corresponding author: Emanuele Viola,
Northeastern University, College of Computer and Information
Science, 360 Huntington Avenue, Boston, MA 02115, (617) 373--8298
(voice), {\tt viola@ccs.neu.edu} (email).}}
\date{This Draft: February 23, 2010}

\maketitle \thispagestyle{empty} \centerline{\large \bf Abstract}
\baselineskip 14pt \vskip 20pt \noindent
We construct a financial ``Turing test'' to determine whether human 
subjects can differentiate between actual vs.~randomized financial 
returns. The experiment consists of an online video-game 
(\texttt{http://arora.ccs.neu.edu}) where players are challenged to 
distinguish actual financial market returns from random temporal 
permutations of those returns.  We find overwhelming statistical evidence 
($p$-values no greater than 0.5\%) that subjects can consistently 
distinguish between the two types of time series, thereby refuting the 
widespread belief that financial markets ``look random''. A key feature of 
the experiment is that subjects are given immediate feedback regarding the 
validity of their choices, allowing them to learn and adapt. We suggest 
that such novel interfaces can harness human capabilities to process and 
extract information from financial data in ways that computers cannot. 


\newpage

\setcounter{page}{1}
\onehalfspacing
\setlength{\belowdisplayskip}{0.250in}
\setlength{\abovedisplayskip}{0.250in}
\setlength{\belowdisplayshortskip}{0.250in}
\setlength{\abovedisplayshortskip}{0.250in} \setcounter{equation}{0}

\section{Introduction}%
\label{sec:intro}%
\par Market efficiency---the idea that ``prices fully reflect all
available information''---is one of the most important concepts in
economics. A vast literature has been devoted to its formulation,
statistical implementation, and refutation since Samuelson (1965)
and Fama (1965a,b and 1970) first argued that price changes, or
returns, must be unforecastable if they fully incorporate the
information and expectations of all market participants. The more
efficient the market, the more random the sequence of returns
generated by it, and the most efficient market of all is one in
which returns are completely random and unpredictable.

\par
Although there is compelling statistical evidence that financial security
prices do not always follow random walks (Lo and MacKinlay, 1988 and 1999), the
belief that human beings cannot distinguish market returns from randomly
generated ones is widespread. For example, Malkiel (1973) discusses
an experiment in which students were asked to generate returns by
tossing fair coins, which yielded observations that were apparently
indistinguishable from market returns (p.~143). Along the same
lines, Kroll, Levy, and Rapoport (1988) conduct an experiment of a
portfolio selection problem, where 40 subjects are asked to choose
between two assets whose returns are sampled randomly and
independently from normal distributions, and given the option of
viewing the assets' past return series.  The authors find that
``even in the extreme case of our experiment, where the subjects
were instructed and could actually verify that the stock price
changes were random, many of them still developed, maintained for a
while, discarded, and generated new hypothesis about nonexistent
trends'' (p.~409).  The same conclusions are reached by De Bondt
(1993), who in a series of experiments about forecasting stock
prices and exchange rates,
finds that ``people are prone to discover `trends' in past prices
and to expect their continuation,'' even when ``stock prices changes
are highly unpredictable''
(p.~357).  Similar experiments are reported in Roberts (1959), Keogh
and Kasetty (2003), and Swedroe (2005), and summarized in Warneryd
(2001).  The belief that humans cannot tell real market data from
random data stands in sharp contrast to those finance practitioners
known as ``technical analysts'' who study past returns with the aim
of forecasting future returns, a task that  is impossible for
randomly generated returns and, therefore, should not be possible
for market returns either. Technical analysts often look for
particular geometric patterns in market returns, e.g., ``head and
shoulders,'' while disciples of efficient markets argue that the
same patterns appear in randomly generated returns, making such
pattern-matching algorithms useless for prediction.

\newpage

\begin{figure}
\begin{center}
\includegraphics[scale=0.8]{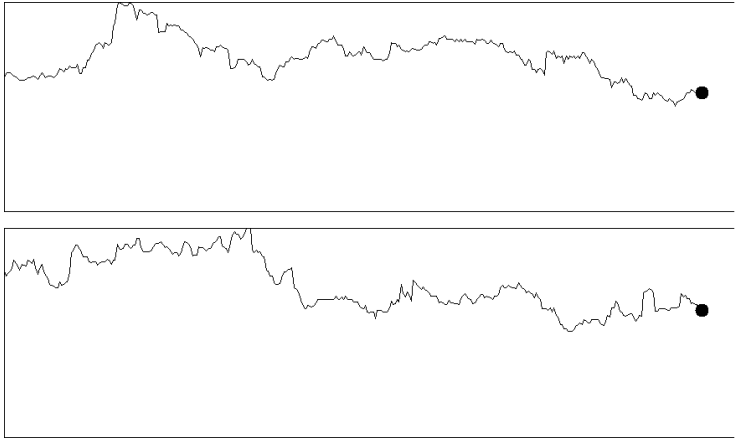}
\end{center}
\caption{Reindeer (real data in top panel).} \label{pic-reindeer}
\end{figure}

\begin{figure}
\begin{center}
\includegraphics[scale=0.8]{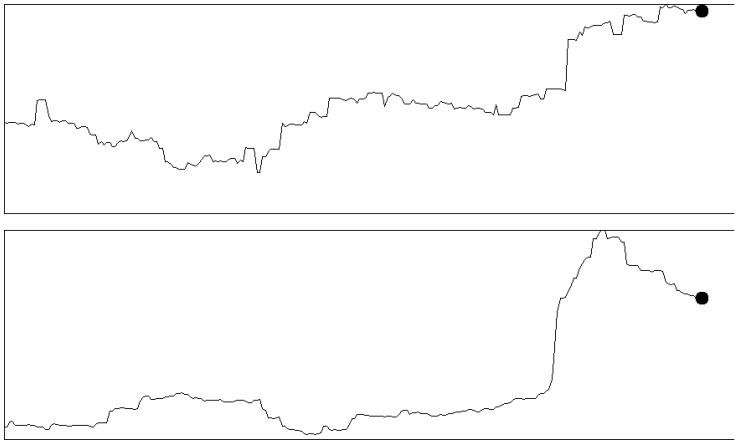}
\end{center}
\caption{Bear (real data in bottom panel).} \label{pic-bear}
\end{figure}

\begin{figure}
\begin{center}
\includegraphics[scale=0.8]{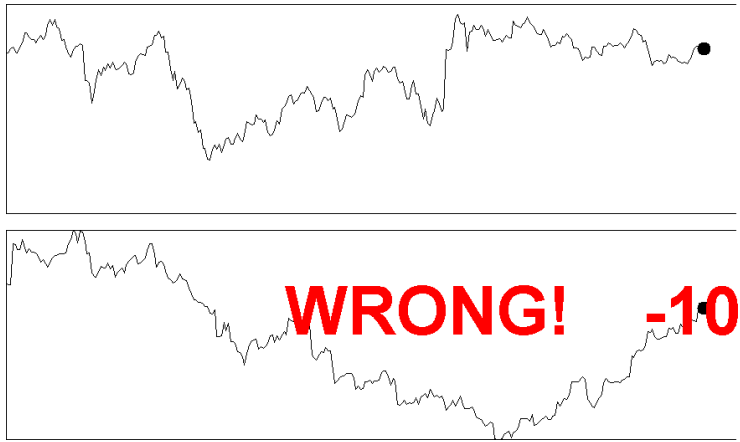}
\end{center}
\caption{Wrong choice in Beaver contest.} \label{pic-beaverwrong}
\end{figure}

\begin{figure}
\begin{center}
\includegraphics[scale=0.8]{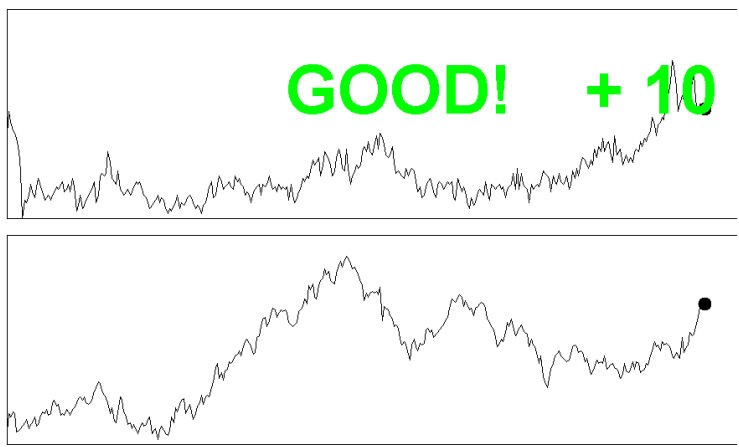}
\end{center}
\caption{Correct choice in Elk contest.} \label{pic-elkgood}
\end{figure}

In this paper we report the results of an experiment designed to test the 
ability of human subjects to distinguish between actual and randomly 
generated returns of financial securities.  We develop a simple web-based 
video-game in which subjects are shown two dynamic price series side by 
side---both of which display price graphs evolving in real time (a new 
price realized each second)---but only 
one of which is a ``replay'' of actual historical price series.%
\footnote{See \texttt{{h}ttp://arora.ccs.neu.edu}.}
The other series is constructed from a random shuffling of the actual 
series, which preserves the marginal distribution of the returns but 
eliminates any time-series properties, effectively creating a random walk 
for prices (Figures \ref{pic-reindeer} and \ref{pic-bear}).  Subjects are 
asked to press a button indicating their selection of the actual price 
series, and are informed immediately whether they were correct or 
incorrect (Figures \ref{pic-beaverwrong} and \ref{pic-elkgood}), after 
which the next pair of price series begins being displayed. 


In a sample of 78 subjects participating in up to 8 different contests 
(using different types of financial data),%
\footnote{Specifically, 78 accounts were created, each corresponding to a 
unique e-mail address.} 
with each contest lasting two weeks and concluding with prizes awarded to 
top performers, we obtained 8015 human-generated guesses for this 
real-time choice problem.  The results provide overwhelming statistical 
evidence ($p$-values of at most $0.5\%$) that humans can quickly learn to 
distinguish actual price series from randomly generated ones. 

We put forward our experiment as a kind of financial ``Turing test.''
As in Turing's original formulation
(Turing, 1950)---a computer passes his test if a
human subject cannot distinguish between interactions with it and another
human subject---our experiment is meant to determine whether humans can
distinguish actual financial data from randomly generated data.
Interaction is a key component of both tests; in our case it amounts to
human subjects receiving feedback about their guesses. To date, no known
computer has passed Turing's test. Similarly, our findings indicate that
as of now, financial markets have not passed our financial Turing test.


\section{Experiment Design}%
\label{sec:experiment}
To test the null hypothesis ${\rm H}$ that human subjects cannot 
distinguish between actual and randomly generated price series, we begin 
with a time series of actual historical prices $\{p_0, p_1, p_2,\ldots, 
p_T\}$ and compute the returns or price differences $\{r_t\}$, 
\begin{equation}
r_t \ \ \equiv\ \ p_t\ -\ p_{t-1}
\end{equation}
from which we construct a randomly generated price series $\{p^*_0, p^*_2, 
\ldots, p^*_T\}$ by cumulating randomly permuted returns: 
\begin{equation}
p^*_t\ \ \equiv\ \ \sum_{k=1}^t r_{\pi(k)}~~~,~~~p^*_0\ \equiv\ p_0 ~~~,~~~
\pi(k)\ :\ \{1,\ldots,T\}\ \to\ \{1,\ldots,T\}
\end{equation}
where $\pi(k)$ is a uniform permutation of the set of time indexes 
$\{1,\ldots,T\}$. A random permutation of the actual returns does not 
alter the marginal distribution of the returns, but it does destroy the 
time-series structure of the original series, including any temporal 
patterns contained in the data.  Therefore, the randomly permuted returns 
will have the same mean, standard deviation, and moments of higher order 
as the actual return series, but will not contain any time-series patterns 
that can be used for prediction.  This construction will allow us to test 
specifically for the ability of human subjects to engage in visual pattern 
recognition in financial data. 

\begin{figure}[htbp]
\begin{center}
\includegraphics[scale=0.6,clip,trim=0 230 0 -160]{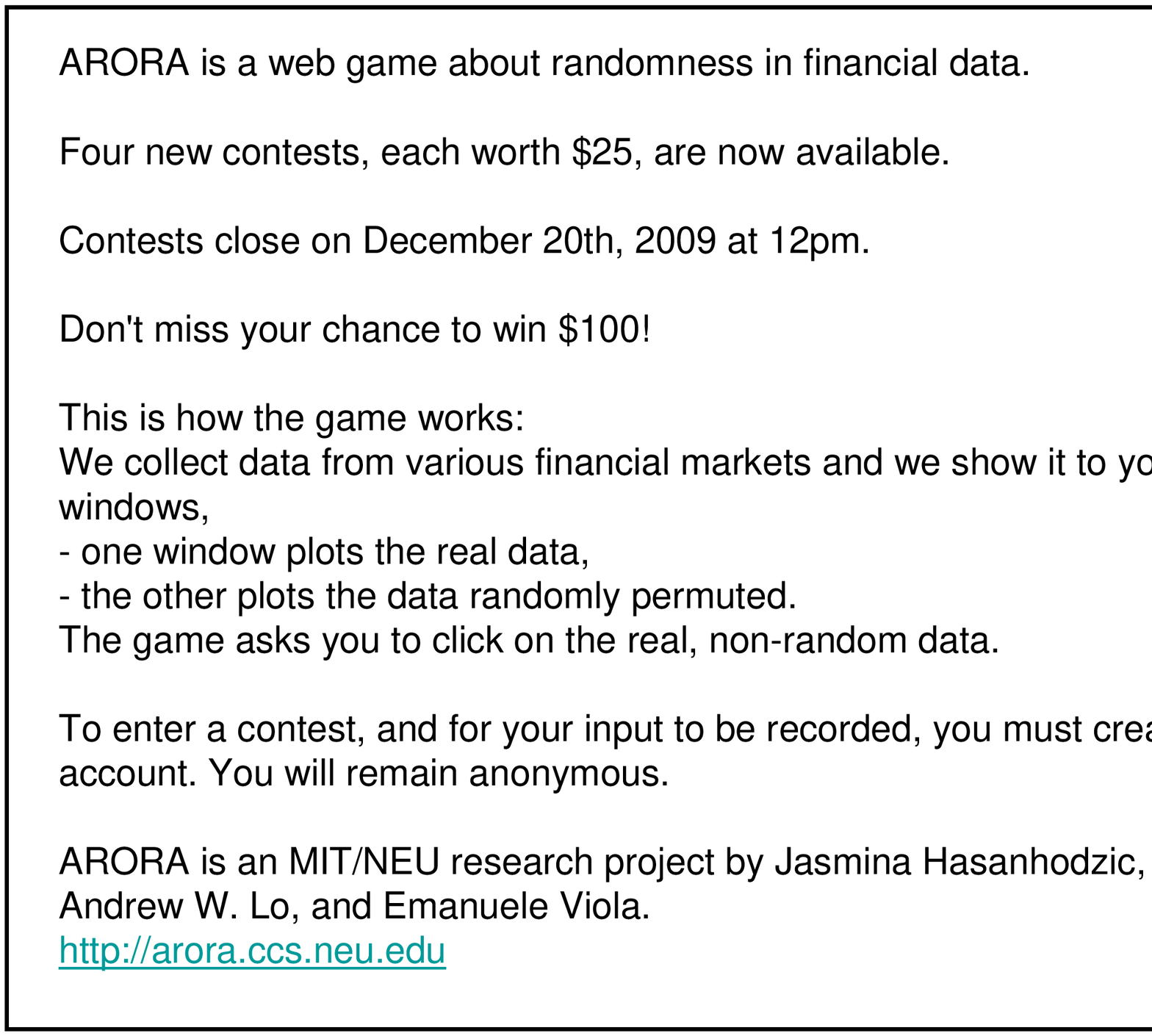}
\end{center}
\caption{Call for subjects for ARORA.}
\label{fig:ad}
\end{figure}

To implement this comparison, we developed a web-based video-game which 
was advertised via email and on websites to computer science students at 
Northeastern University, MBA students at the MIT Sloan School of 
Management, and finance practitioners 
(see Figure \ref{fig:ad} for a sample advertisement).%
\footnote{The video-game can be accessed at 
\texttt{http://arora.ccs.neu.edu}. To recruit subjects, an announcement 
was emailed to Northeastern computer science students, MIT Sloan MBA 
students in the Fall section of 15.970, members of the Market Technicians 
Association mailing list, the MTA Educational Foundation mailing list, the 
American Association of Individual Investors mailing list, and the staff 
and Twitter followers of TraderPsyches.} 
After registration, a subject can participate in trials from eight 
different contests, each consisting of the same game applied to different 
data sets. The data sets consist of returns of eight commonly traded 
financial assets: the NASDAQ Composite Index, the Russell 2000 Index, the 
US Dollar Index, Gold (spot price), the Dow Jones Corporate Bond Price 
Index, the Dow Jones Industrial Average, the Canada/US Dollar Foreign 
Exchange Rate, and the S\&P GSCI Corn Index (spot 
price).%
\footnote{The Dow Jones Corporate Bond Price Index was obtained 
from the Global Financial Database, while all other data series were 
obtained from Bloomberg.}
These data sets were arbitrarily named after 
animals, so that users had no knowledge of the specific financial assets 
used in the experiment. 

Participating in a trial consists of the following task. The subject
is shown two dynamic price charts on a computer screen, one above
the other (Figures \ref{pic-reindeer} and \ref{pic-bear}). Each
graph evolves through time---similar to those appearing in computer
trading platforms---plotting the price at that point in time as well
as the trailing prices over a fixed time window over the most recent
past. Prices are defined as the cumulative sum of a sequence of
returns.  Of the two moving charts, only one corresponds to the
sequence of market returns from the actual data set; we call this
graph the ``real'' chart or $\{p_t\}$. The other corresponds to the
sequence of returns obtained by randomly permuting the sequence of
market returns; we call this graph the ``random chart'' or
$\{p^*_t\}$.  The computer chooses at random which of the two graphs
is placed at the top or the bottom.

The subject is asked to decide which of the two moving charts is the
real one by clicking on it. The game registers the subject's choice,
and informs the subject immediately whether his/her guess is correct
or incorrect (Figures \ref{pic-beaverwrong} and \ref{pic-elkgood}).
For each data set, the user is shown approximately $35$ pairs of
moving charts and asked to make as many choices. The subject is also
free to refrain from choosing. This happened rarely, and to err on
the conservative side, we recorded the absence of a guess as an
incorrect choice for that trial.\footnote{For two of the data sets,
we also dropped one subject from each because the two subjects
provided responses for less than 50\% of the trials.} To provide
the participants with some incentive for making correct choices,
top-scoring players were awarded prizes (\$10 or \$25 Amazon gift
certificates).

To evaluate the robustness of our experimental design, we varied
various parameters of the experiment across data sets, as indicated
in the Results section below. In addition, we presented subjects
with data charts in two different ways. For half of the data sets
corresponding to transaction-by-transaction (or ``tick'') data, each
subject was shown a fresh set of charts, based on a sequence of
returns disjoint from the sequences shown to other subjects. For the
other half of the data, corresponding to daily data, the charts
shown to each subject were based on the same sequence of
returns.\footnote{However, the data was shifted by a random amount
for security reasons, i.e., to avoid the possibility that two
subjects could coordinate their guesses, for example by
simultaneously playing the same charts on two nearby machines.}

Finally, for each data set, subjects were offered the opportunity to
practice on a separate set of data.

\begin{figure}[htbp]
\begin{center}
\includegraphics[scale=0.88]{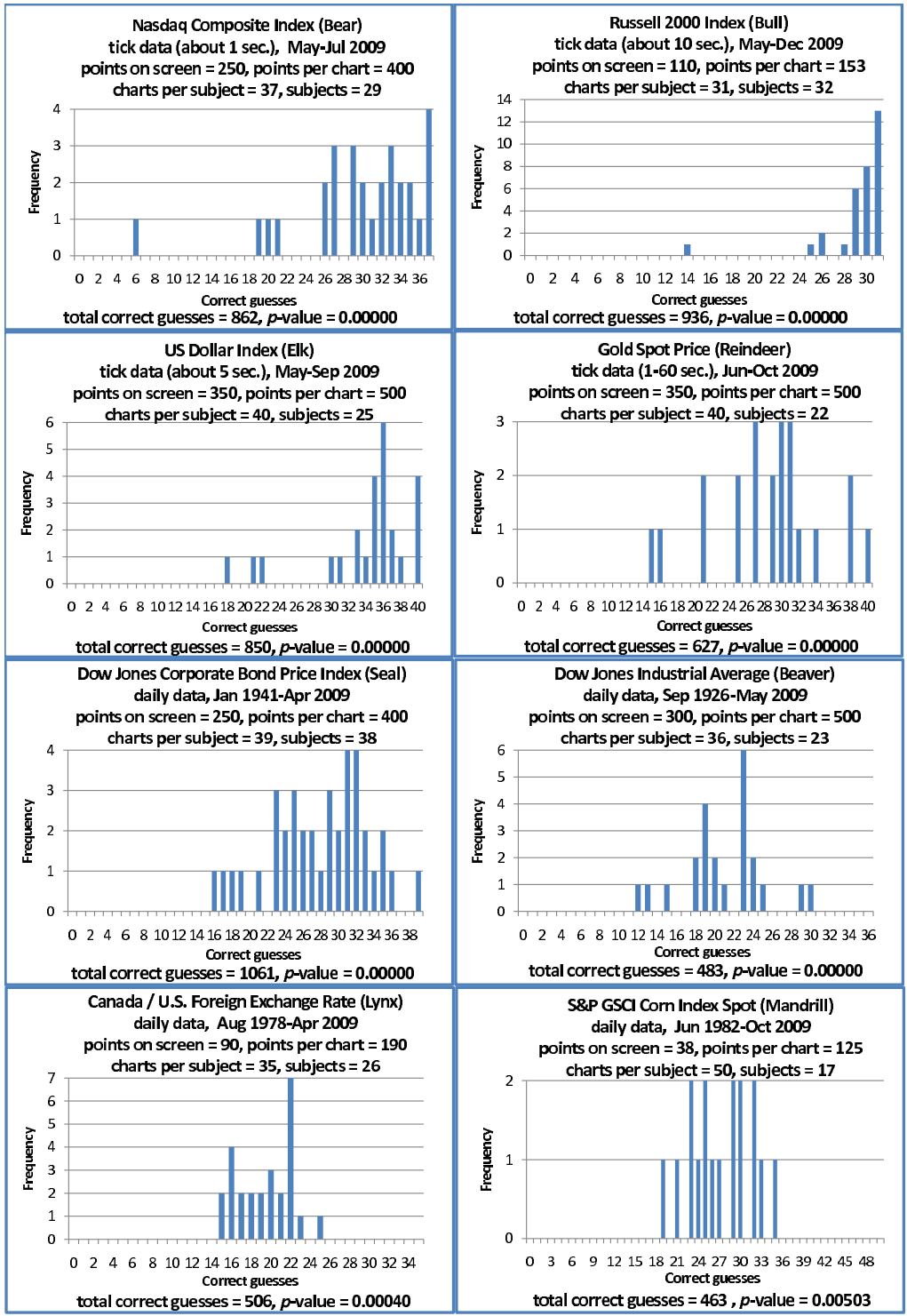}
\end{center}
\caption{Summary of experimental results across all eight contests.} \label{pic-results}
\end{figure}

\section{Results}%
\label{sec:results}
The results are summarized in Figure \ref{pic-results}. For each
data set we report how many return observations are presented to
each subject (points per chart). As the charts are moving, we also
report how many returns are present on the screen at any moment
(points per screen). We then report how many pairs of charts each
subject was presented with (charts per subject), and how many
subjects participated (subjects). The distribution of correct
guesses across subjects is reported in a histogram, and the sum of
correct guesses across all subjects is below that.

Under the null hypothesis ${\rm H}$, human subjects should not be
able to distinguish between real and random charts, so their choices
should be no better than purely random guesses.  Therefore, testing
the null hypothesis involves computing the probability value, or
$p$-value, of obtaining at least as many correct guesses when
guessing at random, i.e., by tossing a fair coin. Specifically, for
a given data set where $s$ subjects were shown $c$ charts each,
suppose the experiment resulted in a total of $g$ correct guesses.
The $p$-value is computed as the probability that the number $X$ of
``heads'' in $n \equiv s \cdot c$ independent tosses of a fair coin
is at least $g$:
\begin{equation}
\mbox{$p$-value}\ \ \equiv \Pr[X \ge g]\ \ =
\ \ \sum_{i = g}^{n} \binom{n}{i}/2^{n}\ .
\end{equation}
For example, the data set ``Lynx'' consists of $s = 26$ subjects
that were shown $c = 35$ charts and made $g = 506$ correct choices,
implying a $p$-value of $0.00040$ or $0.040\%$.

The $p$-values for each data set are reported in Figure
\ref{pic-results}. The evidence against the null hypothesis is
overwhelming: the $p$-values are at most $0.503\%$ for each of the
eight data sets, and are less than $0.001\%$ for six of them.

\section{Discussion and Conclusion}%
\label{sec:conclusion}
A natural question that arises is how the subjects managed to
perform so well. One may wonder whether the eight data sets
presented were selectively chosen from a larger universe of results
based on performance.  In fact, the results presented comprise the
entire experiment. We also considered the possibility of a
potentially biased pool of subjects; perhaps those with greater
familiarity with financial data were drawn to this challenge.
Subjects were asked to specify their profession, and the percentage
of correct guesses for those 23 out of 78 who declared finance as
their main profession is virtually indistinguishable from that of
the others (73.6\% for finance professionals vs.\ 72.2\% for the
others).  In our experiment, financial experience seems to have no
correlation with performance.  Skeptics may wish to try the
challenge for
themselves, and
demonstrate that in short order, they can become quite skilled at
differentiating real financial data from randomized series.

Instead, we conjecture that feedback---which allows subjects to
learn and adapt---is the most significant factor in allowing typical
subjects to distinguish real market returns from their randomized
counterpart.  Casual inspection of Figures
\ref{pic-reindeer}--\ref{pic-elkgood} shows that distinguishing real
data from randomized data is challenging; for some data sets the
real chart tends to be smoother, as in Figure \ref{pic-bear}, while
for other data sets the opposite is true, the real chart tends to be
spikier, as in Figure \ref{pic-elkgood}. But for each data set,
feedback from just a few trials seems sufficient for the user to
extract characteristics of the data to be used in classifying future
charts. Our conjecture is supported by the information about winning
strategies that some of the subjects volunteered to share with us
(anonymously). For example, a subject wrote:
\begin{quote}
Admittedly, when first viewing the two data sets in the practice
mode, it is impossible to tell which one is real, and which one is
random, however, there is a pattern that quickly emerges and then
the game becomes simple and the human eye can easily pick out the
real array (often in under 1 second of time). In the Bull and Bear
games, the real data array was smoother and less volatile, while for
the Elk and Reindeer games it was the opposite: the real array was
more noisy.
\end{quote}
The human eye---as opposed to a computer algorithm---may have a
crucial advantage. It is well known that computers still struggle
with many image-recognition and classification tasks that are
trivial for humans, and the same may be said for distinguishing
market returns from randomized versions.  This gap between human and
algorithmic pattern recognition may explain the gulf separating
technical analysis (a largely human endeavor) and quantitative
financial analysis (a more analytical and algorithmic approach), and
why the former practice persists despite the lack of support from
the latter.

More generally, human intelligence is intertwined with pattern recognition
and prediction (Hawkins, 2004), and financial pattern recognition is just
one of many domains in which we excel.  Our simple experimental framework
suggests the possibility of developing human/computer interfaces that
allow us to translate certain human abilities into other domains and
functional specifications.  For example, with the proper interface, it may
be possible to translate the hand-eye coordination of highly skilled
video-gamers to completely unrelated pattern-recognition and prediction
problems such as weather forecasting or financial trading.   We hope to
explore such interfaces in future research.

\newpage

\nocite{*}
\bibliography{alphassrn}
\end{document}